\documentclass[12pt,preprint,super,squares,numbers]{aastex}

\def\teff{$\rm T_{\rm eff}$}
\def\alfa{$\rm [\alpha/Fe]$}

\def\ltsima{$\; \buildrel < \over \sim \;$}
\def\gtsima{$\; \buildrel > \over \sim \;$}
\def\lsim{\lower.5ex\hbox{\ltsima}}
\def\gsim{\lower.5ex\hbox{\gtsima}}
\def\lapp{\ifmmode\stackrel{<}{_{\sim}}\else$\stackrel{<}{_{\sim}}$\fi}
\def\gapp{\ifmmode\stackrel{>}{_{\sim}}\else$\stackrel{<}{_{\sim}}$\fi}

\usepackage{amsmath}
\usepackage{textcomp}
\usepackage{amssymb}
\usepackage{multirow}
\usepackage{booktabs}
\usepackage{graphicx}
\usepackage[export]{adjustbox}
\usepackage{subfigure}
\usepackage{lscape}
\usepackage{longtable}

\usepackage[superscript,biblabel]{cite}

\newsavebox{\bigimage}

\newdimen\minuswidth    
\setbox0=\hbox{$-$}
\minuswidth=\wd0
\catcode`@=\active
\def@{\kern\minuswidth}
\setbox0=\hbox{\rm0}

\begin{document}
\title{A relic from a past merger event in the Large Magellanic Cloud}

\author{{\rm
A. Mucciarelli\altaffilmark{1,2},
D. Massari\altaffilmark{2,3},
A. Minelli\altaffilmark{1,2},
D. Romano\altaffilmark{2}, 
M. Bellazzini\altaffilmark{2}, \\
F. R. Ferraro\altaffilmark{1,2},
F. Matteucci\altaffilmark{4,5,6},
L. Origlia\altaffilmark{2}
}}

\affil{\altaffilmark{1}{\rm \it \small Dipartimento di Fisica e Astronomia, Universit\`a degli Studi di Bologna, via Gobetti 93/2, I-40129 Bologna, Italy}}
\affil{\altaffilmark{2}{\rm \it  \small INAF - Osservatorio di Astrofisica e Scienza dello Spazio di Bologna, via Gobetti 93/3, I-40129 Bologna, Italy }}
\affil{\altaffilmark{3}{\rm \it  \small 
University of Groningen, Kapteyn Astronomical Institute, NL-9747 AD Groningen, The Netherlands}}
\affil{\altaffilmark{4}{\rm \it  \small Dipartimento di Fisica, Sezione di Astronomia, Universit\`a di Trieste, Via Tiepolo 11, 34131 Trieste, Italy }}
\affil{\altaffilmark{5}{\rm \it  \small INAF, Osservatorio Astronomico di Trieste, Via Tiepolo 11, 34131 Trieste, Italy}}
\affil{\altaffilmark{6}{\rm \it  \small INFN, Sezione di Trieste, Via Valerio 2, 34127 Trieste, Italy}}

{\bf 
According to the standard cosmological scenario, the large galaxies
that we observe today have reached their current  mass via mergers with smaller galaxy 
satellites \citep{moore99}. This hierarchical process is expected to take place 
on smaller scales for the satellites themselves, that should build-up from the 
accretion of smaller building blocks \citep{donghia08}.
The best chance we have to test this prediction is by looking at the most massive satellite 
of the Milky Way (MW): the Large Magellanic Cloud (LMC).
Smaller galaxies have been revealed to orbit around the LMC \citep{erkal20,patel20}, but 
so far the only evidence for mutual interactions is related to the orbital interplay 
with the nearby Small Magellanic Cloud (SMC), which is the most massive LMC satellite.
In this work, we report the likely discovery of a past merger event that the LMC experienced with
a galaxy with a low star formation efficiency and likely
having a stellar mass similar to those of dwarf spheroidal galaxies.
This former LMC satellite has now completely dissolved, 
depositing the old globular cluster (GC) NGC~2005 as part of its debris. 
This GC is the only surviving witness of this ancient merger event, recognizable 
through its peculiar chemical composition. 
This discovery is the observational evidence that the process of 
hierarchical assembly has worked also in shaping our closest satellites.}

The LMC is the largest satellite orbiting the MW, with a total mass of $\sim1-2.5\times10^{11}$ M$_\odot$ 
\citep{penarrubia16,erkal19} and a stellar mass of $\sim3\times10^{9}$ M$_\odot$ \citep{vdm14}.
A satellite this massive is expected to host its own system of satellites. According to
models of galaxy formation in the LCDM theory, the number of these satellites is in the range 4-40 
\citep{guo10,sales13}, the most massive of them dominating the mass budget (with a mass ratio compared 
to the LMC of $\sim0.1$). The SMC, with a total mass of $\sim2\times10^{9}$ M$_\odot$ \citep{stan04}, 
matches well this prediction. The precise measurement of their proper motion allowed for the first time 
a reasonably sound reconstruction of the orbital history of the system\citep{kalli06,kalli13}. 
According to the most recent analyses the MCs may have become bound to each other around $\sim$3 Gyr 
ago and had their last close encounter $\sim$150 Myr ago \citep{patel20}.

The other satellites of the LMC should be much smaller, with total masses from $\sim10^{8}M_{\odot}$ 
down to values typical of the ultra faint dwarf galaxies  \citep[UFDs,][]{simon19} that are 
the lowest-luminosity, oldest, most dark matter-dominated galaxies known so far.
Attempts to determine which of the known UFDs were accreted by the MW together with the LMC hugely 
benefited from the advent of the second data release of the Gaia mission \citep{brown18}, as this 
enabled the possibility to determine their 3D kinematics \citep{helmi18,simon18,kalli18}. 
Dynamical integration of the UFDs orbits led to the conclusion that 4 to 6 of them \citep[depending on the details 
of the modeling,][]{erkal20,patel20} are indeed current satellites of the LMC. However, 
nothing is known about the past population of LMC satellites, that may be already disrupted within the host 
galaxy halo. So far, the only traces of accretion of matter from another galaxy by the LMC 
are associated with the complex interaction with the SMC \citep{donghia16,olsen11}.

Chemical tagging \citep{freeman02} is one of the few techniques that allows us to trace completely dissolved 
satellites, also in absence of any kinematically or spatially coherent relic, identifying  stars and clusters 
that were lost long ago by means of their anomalous chemical composition, in contrast with the environment 
in which they live nowadays. 
However the power of the technique can be strongly hampered by the fact that spotting chemically
anomalous stars in a given galaxy requires (a) high resolution spectroscopy for large samples, 
and (b) extremely homogeneous chemical abundance analysis, as subtle differences in the assumptions on, 
e.g., astrophysical parameters, can wipe out (or spuriously introduce) the small abundance differences 
we are looking for. With the aim of digging into the past merging history of the largest MW satellite, 
here we attempt to overcome these problems by using old GCs as tracers and 
by deriving  chemical abundances  from high resolution spectra with a strictly homogeneous analysis.

In this respect, GCs are a class of tracers that has been proven to be particularly effective in reconstructing
the merger history of a galaxy such as the MW \citep{massari19,myeong19,krui20} or M31 \citep{mackey19}.
This is because even a very low mass, low surface brightness dwarf galaxy, that may be dissolved by the tidal
force of the main galaxy at its first peri-galactic passage, may host a dense stellar cluster able to survive 
in the same tidal field for many Gyr. Such a cluster will keep record of the characteristics of the environment
in which it was born. In particular the chemical abundance pattern of its stars may be quite different from 
that of stars and clusters born in the main galaxy, 
due to the large differences in the star formation and chemical evolution between the hosting and 
the progenitor system.
We analyzed optical, high-resolution spectra of red giant stars 
in 11 old LMC GCs and in a reference sample of 15 MW GCs. These two datasets have been analyzed 
with the same methodology (i.e. atomic data, solar reference abundances, model atmospheres, temperature scale),
thus removing any possible systematic error between the abundances of the two families of clusters.

We derived the chemical abundance ratios for 13 species, indicators of different production mechanisms 
and stellar progenitors. 
The LMC GCs draw well-defined sequences of each abundance ratio as a function of [Fe/H] 
that are distinct, in most cases, from those defined by the MW GCs, reflecting the different 
chemical evolution histories of the two galaxies \citep{lapenna12,swae13,nidever20}.
Among the LMC GCs, the metal-poor cluster NGC~2005 ([Fe/H]=--1.75$\pm$0.04 dex) is distinguished as a 
clear outlier. NGC~2005 is a relatively massive GC, 
M$\sim2-3~10^{5} {\rm M_{\odot}}$ \citep{mackey03},
located at $\sim{\rm0.23}$ kpc from the center of LMC.
It exhibits abundance ratios that are systematically lower (in most cases at a level $>$3$\sigma$) 
than those measured in the LMC GCs with similar metallicities (Figure 1) for almost all the species, 
including elements  (Si, Ca, Sc, Ti, V, Mn, Co, Ni, Cu, Zn, Ba, La, Eu)  forming from different nucleosynthesis channels 
(i.e. explosive and thermonuclear SNe, hypernovae, slow and rapid neutron capture processes). 
The 5 LMC GCs with [Fe/H] comparable with that of NGC~2005 (--1.75 $<$[Fe/H]$<$--1.69 dex) have abundance ratios 
very similar each other, constituting a homogeneous group of clusters sharing the same chemistry. 
This demonstrates that these GCs formed in environments that have experienced a similar chemical enrichment history, 
likely the LMC itself. On the other hand, the strong chemical differences between NGC~2005 and this group of clusters 
is indicative of a completely different chemical enrichment path. 
This reveals that  NGC~2005 cannot have formed in the same environment as the rest of the LMC clusters 
at that metallicity but it has rather born in a system that converted its gas into stars at a slower pace.

In order to determine the characteristics of NGC~2005 most likely progenitor, we computed chemical 
evolution models for different galactic environments. The analyzed data allowed us to produce models 
and calibrate them with respect to MW-like and LMC-like environment, and for systems evolving with less efficient 
star formations (see Methods).
We purposely focused on elements with highly accurate stellar yields and that are representative of different 
nucleosynthesis channels (Table 1 and Figure 2): 
Si and Ca (mainly produced through $\alpha$-capture processes), 
Cu (mainly produced through slow neutron-capture processes) and 
Zn (mainly built in hypernovae high-energy explosions). 

Our models for the LMC reproduce reasonably well the data 
for all the LMC GCs but NGC~2005, which is always under-abundant at fixed metallicity. 
We run several chemical evolutionary models for the putative NGC~2005 parent galaxy. 
The ones that fit best the peculiar chemistry of NGC 2005 unavoidably require systems evolving with very 
low star formation efficiency, e.g., of dwarf spheroidal galaxies \citep{tolstoy09}.
Although it is very difficult to set precise limits to the mass of the progenitor based on the chemistry alone,
the very low [Zn/Fe] abundance measured in NGC~2005 with respect to the other MC GCs suggests it formed from 
a gas poorly enriched from massive stars. 
In fact, in the framework of our models, 
Zn comes mainly from low-metallicity, massive ($>$ 30${\rm M_{\odot}}$) stars exploding as hypernovae \citep{romano10a}. 
If the formation of such massive stars is suppressed, less Zn is formed, this resulting in a lower [Zn/Fe] ratio overall. 
It has been recently shown\citep{yan20a} that the very low star formation rates expected in low-luminosity, 
metal-poor stellar systems lead to a lower upper mass limit for the galaxy-wide initial mass function (IMF). 
Indeed, if we consider star formation rates lower than $\sim{\rm 5\cdot10^{-4} M_{\odot} yr^{-1}}$  
and an upper mass limit of 40 $M_{\odot}$ for the galaxy-wide IMF, a remarkably good fit of the observed [Zn/Fe] ratio 
for NGC~2005 is obtained. All the other abundance ratios are fitted well within their errors under the same premises.
These models for NGC~2005 are to be compared with the model for the LMC GCs  that assumes  
a star formation rate of the order 1-1.5 ${\rm M_{\odot} yr^{-1}}$ 
during the early LMC evolution and a galaxy-wide IMF upper mass limit of 100 ${\rm M_{\odot}}$.

The radial velocity of NGC~2005 is similar to that 
of other clusters in its surroundings, hence any (possible) strong anomaly due to its association with an accreted 
satellite has been washed out after many orbits within the gravitational potential of the LMC. On the other hand, 
its peculiar chemical composition suggests that this cluster originated in a galaxy that formed 
its stars with a much less efficient star formation compared to the LMC. This evidence suggests a low-mass 
galaxy progenitor, as massive as the dwarf spheroidal galaxies currently orbiting the MW or even lighter, 
characterized by low star formation rates \citep{tolstoy09}. 
NGC~2005 is the surviving witness of the ancient merger event leading to the dissolution of its parent galaxy 
into the LMC, the only case known so far identified by its chemical fingerprints in the realm of dwarf galaxies.
Our findings thus supports the 
predictions on the self-similar nature of the process of galaxy formation by the standard cosmology 
on our closest satellite, and opens a new way to investigate the assembly history of 
galaxies beyond the Milky Way via the chemical tagging of their GC systems.

\begin{figure}
 \centering%
\includegraphics[scale=0.3]{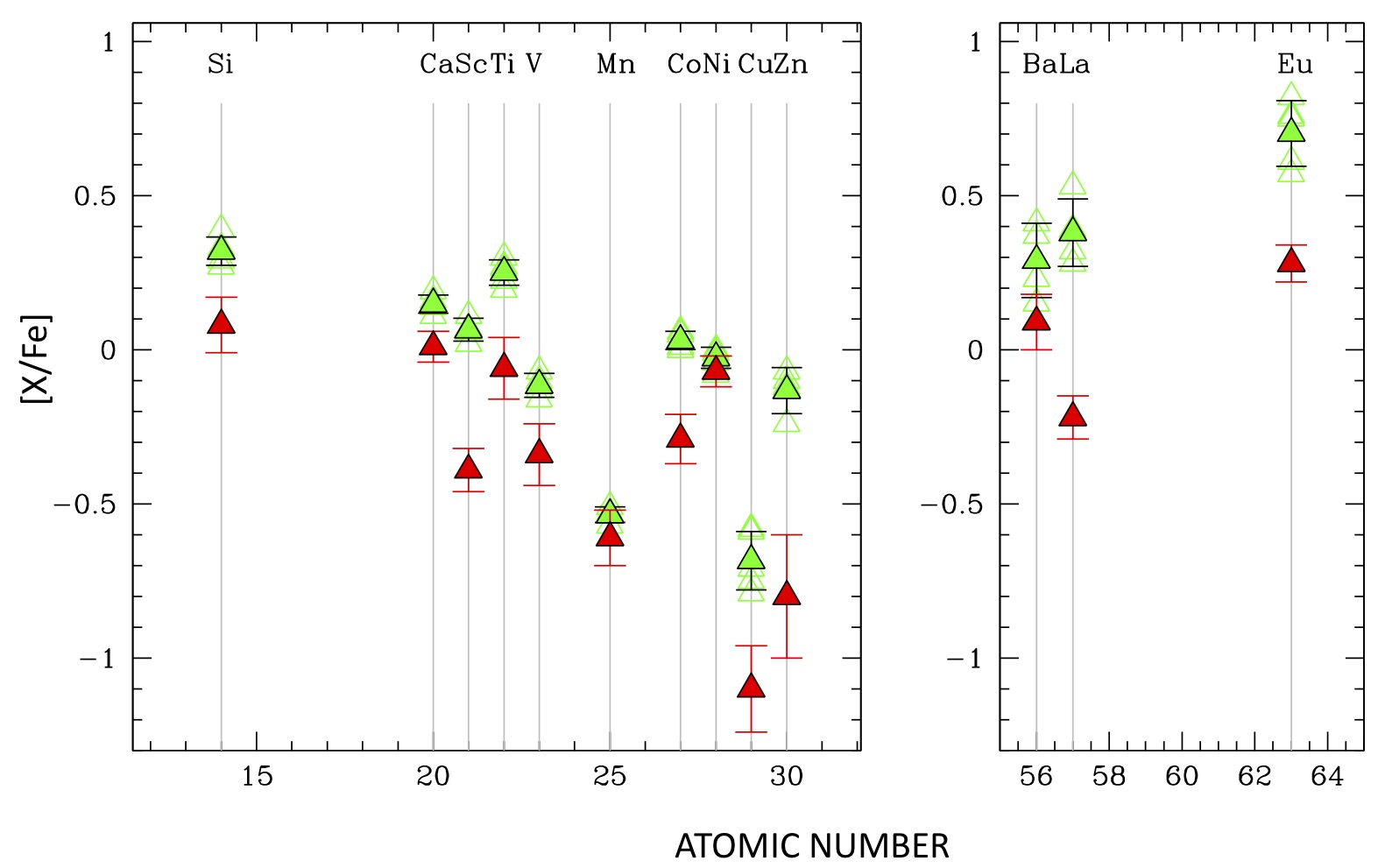}
\caption{
{\bf Chemical abundances of NGC~2005.}
Abundance ratios measured for the accreted cluster NGC~2005 (red triangles) in comparison with those measured 
in the LMC old clusters with comparable metallicity (--1.75$<$[Fe/H]$<$-1.69 dex, green open triangles, namely 
NGC~1786, NGC~1835, NGC~1916, NGC~2210 and NGC~2257). 
The green filled triangles represent the average abundance ratios obtained for these five LMC GCs and the 
errorbars are the corresponding standard deviation.}
\label{metalpoor}
\end{figure}

\begin{figure}
 \centering%
\includegraphics[scale=0.28]{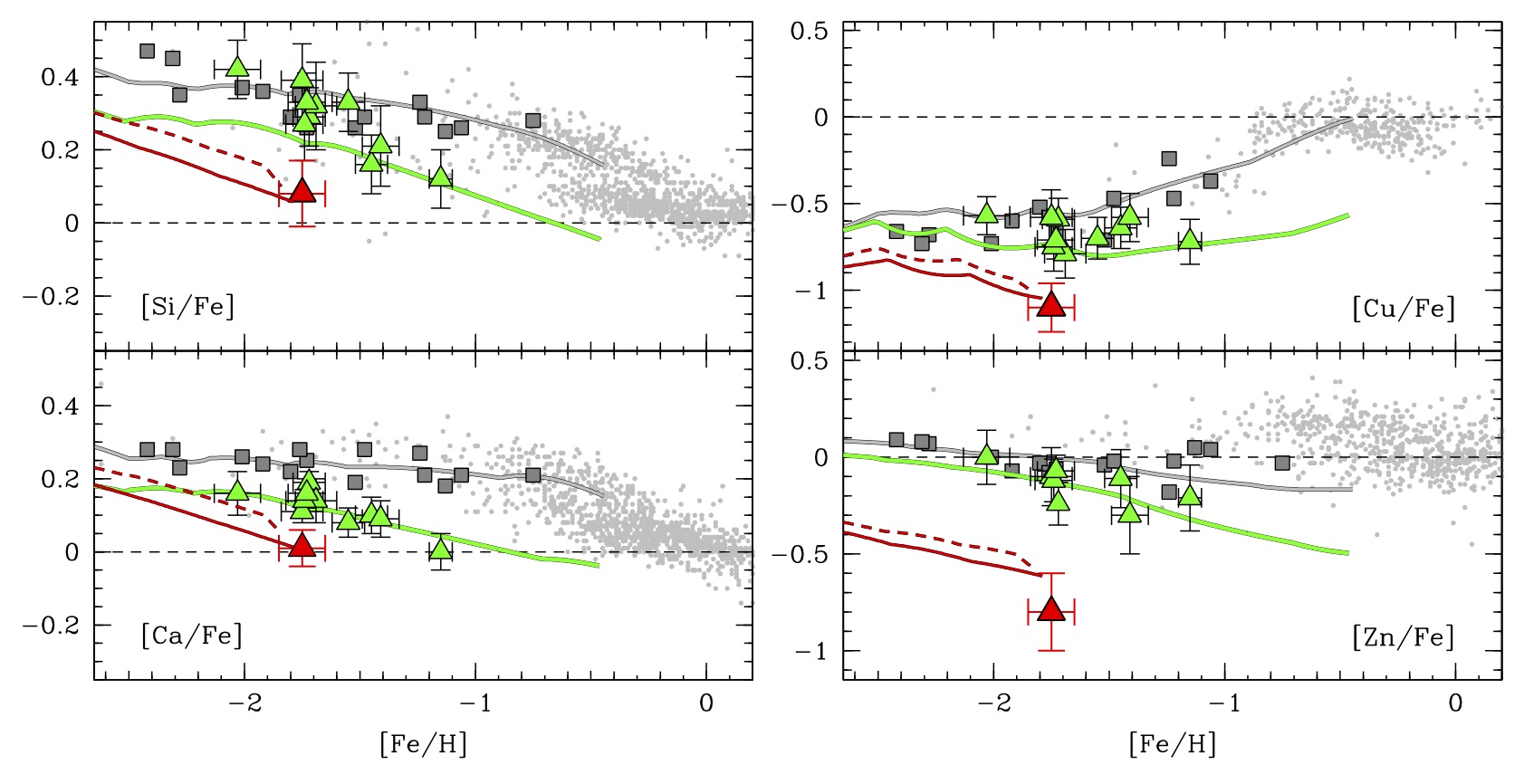}
\caption{
{\bf Chemical abundances of the LMC and MW clusters.}
Behaviour of the [Si/Fe], [Ca/Fe], 
[Cu/Fe] and [Zn/Fe]
abundance ratios as a function of [Fe/H] 
for the LMC (green triangles) and the 
MW (grey squares) clusters.
The accreted LMC cluster NGC~2005 is highlighted as a red triangle.  
Solar neighbourhood stars  \citep[small grey circles,][]{bensby14} are shown 
as reference. Errorbars are computed as the mean 
value of the uncertainties in individual stars and displayed only for the LMC clusters (see Methods).
Superimposed chemical evolution models for the MW Halo (grey line), LMC (green line) and for 
two stellar systems with low star formation efficiencies, namely 
0.075 ${\rm Gy^{-1}}$ over 1 Gyr and 0.15 ${\rm Gy^{-1}}$ over 0.5 Gyr 
(resulting in a star formation rate of $<{\rm 5\cdot10^{-4} M_{\odot}yr^{-1}}$, red solid and dashed lines, respectively).}
\label{trends}
\end{figure}

\tabletypesize{\scriptsize}
\begin{deluxetable}{lccccc}
\tablecolumns{6} 
\tablewidth{0pc}  
\tablecaption{{\bf Chemical abundances of Fe, Si, Ca, Cu and Zn for LMC and MW globular clusters.} Average weighted abundance ratios for [Fe/H], [Si/Fe], [Ca/Fe], [Cu/Fe] and [Zn/Fe] 
for the analyzed LMC and MW old GCs with the corresponding standard error and, in brackets, the dispersion of the weighted mean.}
\tablehead{ 
\colhead{Cluster} & [Fe/H] & [Si/Fe]  & [Ca/Fe] & [Cu/Fe]  & [Zn/Fe] \\
  &   (dex)  &  (dex)  & (dex)  & (dex) & (dex) }
\startdata 
\hline
 NGC~1466       &  --1.55$\pm$0.02 (0.05)    & +0.33$\pm$0.01 (0.02)    &  +0.08$\pm$0.03 (0.08)  & --0.70$\pm$0.11 (0.15) & ---    \\
 NGC~1754       &  --1.45$\pm$0.03 (0.05)    & +0.16$\pm$0.02 (0.04)    &  +0.10$\pm$0.02 (0.04)  & --0.64$\pm$0.07 (0.15) & --0.11$\pm$0.04 (0.10)  \\   
 NGC~1786       &  --1.72$\pm$0.02 (0.04)    & +0.29$\pm$0.05 (0.09)    &  +0.19$\pm$0.03 (0.06)  & --0.59$\pm$0.05 (0.09) & --0.24$\pm$0.05 (0.10)     \\   
 NGC~1835       &  --1.69$\pm$0.01 (0.01)    & +0.32$\pm$0.05 (0.10)    &  +0.14$\pm$0.02 (0.04)  & --0.79$\pm$0.08 (0.15) & ---     \\   
 NGC~1898       &  --1.15$\pm$0.02 (0.05)    & +0.12$\pm$0.01 (0.03)    & +0.00$\pm$0.03 (0.07)  & --0.72$\pm$0.05 (0.10) & --0.21$\pm$0.15 (0.23)     \\  
 NGC~1916       &  --1.75$\pm$0.03 (0.05)    & +0.39$\pm$0.01 (0.02)    &  +0.11$\pm$0.03 (0.04)  & --0.58$\pm$0.05 (0.09) & --0.10$\pm$0.08 (0.11)     \\   
 NGC~2005       &  --1.75$\pm$0.04 (0.06)    & +0.08$\pm$0.01 (0.01)    &  +0.01$\pm$0.03 (0.04)  & --1.10$\pm$0.14 (---) & --0.80$\pm$0.20 (---)     \\   
 NGC~2019       &  --1.41$\pm$0.05 (0.08)    & +0.21$\pm$0.01 (0.01)    &  +0.09$\pm$0.04 (0.08)  & --0.58$\pm$0.03 (0.05) & --0.30$\pm$0.20 (---) \\   
 NGC~2210       &  --1.74$\pm$0.02 (0.06)    & +0.27$\pm$0.03 (0.07)    &  +0.14$\pm$0.01 (0.03)  & --0.75$\pm$0.03 (0.08) & --0.12$\pm$0.07 (0.15)     \\  
 NGC~2257       &  --1.73$\pm$0.02 (0.04)    & +0.33$\pm$0.02 (0.04)    &  +0.16$\pm$0.03 (0.05)  & --0.71$\pm$0.01 (0.01) & --0.07$\pm$0.13 (0.21)     \\   
 HODGE~11       &  --2.03$\pm$0.04 (0.09)    & +0.42$\pm$0.08 (---)     &  +0.16$\pm$0.01 (0.02)  & --0.57$\pm$0.02 (0.03) &  +0.00$\pm$0.05 (0.10)    \\   
\hline 
NGC~104   &   --0.75$\pm$0.01 (0.03)  & +0.28$\pm$0.01 (0.03)     &  +0.21$\pm$0.02 (0.07)   & ---  &  --0.03$\pm$0.03 (0.09)   \\   
NGC~288   &   --1.24$\pm$0.01 (0.04)  & +0.33$\pm$0.01 (0.03)     &  +0.27$\pm$0.01 (0.03)   & --0.24$\pm$0.02 (0.05)  &  --0.18$\pm$0.04 (0.14)   \\   
NGC~1851  &   --1.13$\pm$0.01 (0.04)  & +0.25$\pm$0.01 (0.03)     &  +0.18$\pm$0.01 (0.05)   & ---  &  +0.05$\pm$0.03 (0.14)   \\   
NGC~1904  &   --1.52$\pm$0.01 (0.03)  & +0.26$\pm$0.01 (0.02)     &  +0.19$\pm$0.01 (0.02)   & --0.71$\pm$0.01 (0.04)  & 
--0.04$\pm$0.02 (0.06) \\   
NGC~2808  &   --1.06$\pm$0.02 (0.07)  & +0.26$\pm$0.01 (0.04)     &  +0.21$\pm$0.01 (0.02)   & --0.37$\pm$0.04 (0.12)  &  +0.04$\pm$0.05 (0.17)   \\   
NGC~4590  &   --2.28$\pm$0.01 (0.05)  & +0.35$\pm$0.04 (0.06)     &  +0.23$\pm$0.01 (0.02)   & --0.68$\pm$0.02 (0.04)  &  +0.07$\pm$0.03 (0.10)   \\   
NGC~5634  &   --1.80$\pm$0.02 (0.05)  & +0.29$\pm$0.01 (0.04)     &  +0.22$\pm$0.01 (0.03)   & --0.52$\pm$0.04 (0.11)  &  --0.03$\pm$0.05 (0.15)   \\   
NGC~5824  &   --1.92$\pm$0.02 (0.04)  & +0.36$\pm$0.03 (0.08)     &  +0.24$\pm$0.01 (0.02)   & --0.60$\pm$0.04 (0.11)  &  --0.07$\pm$0.03 (0.07)   \\   
NGC~5904  &   --1.22$\pm$0.01 (0.03)  & +0.29$\pm$0.01 (0.03)     &  +0.21$\pm$0.01 (0.03)   & --0.47$\pm$0.02 (0.06)  &  --0.02$\pm$0.02 (0.09)   \\   
NGC~6093  &   --1.76$\pm$0.01 (0.03)  & +0.35$\pm$0.01 (0.04)     &  +0.28$\pm$0.01 (0.03)   & --0.58$\pm$0.01 (0.03)  &  --0.08$\pm$0.02 (0.07)   \\   
NGC~6397  &   --2.01$\pm$0.01 (0.03)  & +0.37$\pm$0.02 (0.08)     &  +0.26$\pm$0.01 (0.03)   & --0.73$\pm$0.04 (0.09)  &  +0.00$\pm$0.02 (0.06)   \\   
NGC~6752  &   --1.48$\pm$0.01 (0.03)  & +0.29$\pm$0.01 (0.03)     &  +0.28$\pm$0.01 (0.02)   & --0.47$\pm$0.01 (0.06)  &  --0.02$\pm$0.03 (0.12)   \\   
NGC~6809  &   --1.73$\pm$0.01 (0.03)  & +0.26$\pm$0.01 (0.04)     &  +0.25$\pm$0.01 (0.03)   & --0.66$\pm$0.01 (0.05)  &  --0.06$\pm$0.01 (0.05)   \\   
NGC~7078  &   --2.42$\pm$0.02 (0.07)  & +0.47$\pm$0.04 (0.09)     &  +0.28$\pm$0.01 (0.02)   & --0.66$\pm$0.03 (0.07)  &  +0.09$\pm$0.03 (0.12)   \\   
NGC~7099  &   --2.31$\pm$0.01 (0.05)  & +0.45$\pm$0.01 (0.02)     &  +0.28$\pm$0.01 (0.03)   & --0.73$\pm$0.03 (0.10)  &  +0.08$\pm$0.02 (0.08)   \\   

\enddata 
\end{deluxetable}

\newpage

\appendix

\begin{center}
{\bf Methods}
\end{center}

{\bf Spectroscopic datasets}\\
The LMC hosts the largest system of old GCs among the MW satellites, including 13 GCs \citep{ols96} 
with ages comparable to those of the MW \citep{brocato96,olsen98,wk17}. 
Chemical abundances of old LMC GCs based on high-resolution spectra of individual giant stars are available 
for about an half of the entire population \citep{hill00,johnson06,mu10,mateluna12}.
These analyses are based on different methods and assumptions making the comparison 
among the LMC clusters and between MW and LMC clusters affected by several systematics 
(i.e. atomic data, solar reference abundances, model atmospheres, temperature scale...).
In order to highlight similarities and differences in the chemical composition of giant stars in LMC and MW GCs, 
we homogeneously analyzed two samples of high-resolution, optical spectra.

{\sl 1. LMC GCs dataset ---}
This sample includes 11 out of 15 old LMC clusters, four of them (NGC~1466, NGC~1754, NGC~1835, NGC~1916) 
have never been analyzed before using high-resolution 
spectroscopy of individual stars (see Table 2). The dataset is composed of proprietary and archival data collected with 
the spectrographs FLAMES \citep{pasquini02} and UVES \citep{dekker00} at the Very Large Telescope 
of the European Southern Observatory and with the spectrograph MIKE \citep{bernstein03} at the 
Magellan Telescope. Signal-to-noise ratios per pixel range from about 30-40 to 100.
For nine GCs, observations with the fiber-fed spectrograph FLAMES in the UVES+GIRAFFE combined mode have been secured. 
For all these clusters spectra with the Red Arm 580 UVES setup have been obtained, 
with a spectral resolution of 47000 and a spectral coverage between about 4800 and 6800 \AA . 
Only for the clusters NGC~1466, NGC~1786, NGC~1898 and NGC~2257, a few of additional cluster stars 
have been observed with the GIRAFFE fibers. In fact, the 
small angular size (about 2 arcmin of diameter) of the LMC clusters and the physical size 
of the magnetic buttons sustaining the fibers prevent to allocate more than $\sim$8-10 FLAMES fibers on the cluster 
area in the same pointing. The adopted GIRAFFE/MEDUSA setups are  HR11 (5597 - 5840 \AA\ and resolution 29500) 
and HR13 (6120 - 6405 \AA\ and resolution 26400).

For two clusters observed with FLAMES (namely, NGC~2210 and NGC~2257), additional archival data acquired with the slit 
spectrograph UVES are available. These observations have been secured with the Red Arm 580 UVES setup, 
adopting slits between 1 and 1.2 arcsec, providing spectral resolutions between 38,000 and 45,000.

Finally, we analyzed MIKE spectra for four GCs 
\citep[NGC~1898, NGC~2005, NGC~2019, Hodge 11,][]{johnson06}, 
two of them in common with FLAMES. The MIKE spectra have been acquired with a slit of 1 arcsec, corresponding 
to a spectral resolution of 19,000 and with a spectral range between 4500 and 7250 \AA\ . 

Among the LMC GCs with metallicity between --1.75 and --1.69 dex (see Figure 1), NGC~2005 is the only
for which the spectra have been obtained with the spectrograph MIKE while 
the other 5 GCs have been observed with the spectrograph UVES. 
Below we discuss the consistency between the abundances obtained from UVES and MIKE spectra.

Note that the previous analysis of NGC~2005 \citep{johnson06} provides abundances consistent with our ones 
but the lack of other LMC GCs with comparable metallicity in that sample did not allow to highlight the peculiarity of the cluster.

{\sl 2. MW GCs dataset ---}
A sample of giant stars in 15 MW GCs has been collected from archival data (see Table 3). 
The clusters have been selected in order to cover the entire range of metallicity 
of the Galactic halo/disk GCs system ([Fe/H] between --2.5 dex and --0.7 dex). 
All the spectra have been obtained with the multi-object spectrograph UVES-FLAMES 
adopting the same setup used for the LMC clusters. 
Signal-to-noise ratios per pixel range from about 70-80 to 150.

{\bf Chemical analysis tools}\\
The chemical abundances of Fe, Si, Ca, Ti and Ni have been derived by comparing the 
measured equivalent widths, derived with the code DAOSPEC \citep{stetson08},  with 
the theoretical line strengths using the code GALA \citep{mu2013}. 
For these species, we considered only transitions selected to be unblended 
according to the atmospheric parameters and metallicity of each individual star, privileging, when possible, 
the lines for which laboratory oscillator strengths are available.\\
Abundances of Sc, V, Mn, Co, Cu, Ba, La and Eu (whose transitions are affected by hyperfine/isotopic splitting) 
and of Zn have been derived through spectral synthesis, by performing a $\chi^2$-minimization between observed 
and synthetic spectra. 
In the case of Zn, the only available line at 4810 \AA\ is not affected by particular splitting or blending. 
However, we resort to spectral synthesis in order to have a better control on the continuum placement, 
which requires particular care because the Zn line is located in a crowded and noisy spectral region. 
This is especially true for some LMC clusters observed with UVES-FLAMES,
that have a lower SNR at the shortest wavelength due to the curve of efficiency of the spectrograph.

All synthetic spectra used in this work have been computed 
with the code SYNTHE \citep{kur05} including all the atomic and molecular transitions available in the Kurucz/Castelli database.\\
Model atmospheres for each star have been calculated with the code ATLAS9 \citep{kur05} under 
the assumptions of plane-parallel geometry, hydrostatic and radiative equilibrium and local 
thermodynamic equilibrium for all the species. 
For all the stars the model atmospheres have been computed assuming an $\alpha$-enhanced chemical mixture, 
except for the stars of NGC~2005 and NGC~1898, for which solar-scaled model atmospheres have been used in accordance 
to the derived [$\alpha$/Fe] abundance ratios. Still, we also verified that the use of $\alpha$-enhanced model atmospheres 
in these two cases changes the measured abundance ratios only slightly, by less than 0.05 dex. 
This means, in particular, that the observed differences between NGC~2005 and the other LMC GCs of similar metallicity 
cannot be reconciled by the adoption of different model atmospheres.

Note that we exclude from this discussion the light elements Na, O, Mg and Al because they are involved in the chemical 
anomalies due to the self-enrichment processes that characterized the early stage of life of the clusters \citep{c09,mu09}.
Therefore, their abundances cannot be easily used as tracers of the chemical composition of the parent galaxy.
Indeed, we found evidence of star-to-star variations for the light elements Na, O, Mg and Al in the target clusters, 
as expected considering their mass and age. 
On the other hand, a null spread has been found for all the elements discussed in this work, 
so that any effect due to the internal evolution of the individual clusters does not affect our conclusions. 

{\bf Determination of the atmospheric parameters}\\
The effective temperature (\teff\ ) is the most crucial atmospheric parameter in the 
determination of chemical abundances. Temperatures can be inferred 
from suitable calibrations of broad-band colors or by requiring that no trend exists between 
the abundances of individual iron lines and their excitation potential. 
The two methods can often provide discrepant results. 
In particular, the two approaches 
agree with each other for metallicities higher than --1.5 dex while the spectroscopic \teff\ are overly low and under-estimated
(down to about 300 K) for [Fe/H]$<$--1.5 dex, because of the 
inadequacies in the modeling of 1D/LTE radiative transfer in metal-poor giant stars \citep{mb20}. 
Therefore, the use of spectroscopic \teff\ leads to underestimate the abundances for metal-poor stars.

Due to the composite nature of the LMC dataset, homogeneous photometric information are not available 
for all the targets: for the proprietary data, near-infrared JH$K_s$ photometry is available, while 
for the archival data, optical ground-based or space-telescope photometry is in hand but in different 
photometric filters. 
Thanks to the high spectral resolution, the high number of lines and the good/high signal-to-noise 
ratio of the LMC spectra, \teff\ can be derived spectroscopically with high precision for all the targets. 
Because the discrepancy between spectroscopic and photometric \teff\ for clusters with [Fe/H]$<$--1.5 dex increases 
with decreasing the metallicity, we need to remove this effect in order to put all the \teff\ on the same (unbiased) scale.
The spectroscopic \teff\ for clusters with [Fe/H]$<$--1.5 dex have been 
corrected according to the spectroscopic [Fe/H] \citep{mb20} in order to put them onto a photometric scale \citep{gh09}, 
while spectroscopic \teff\ for clusters with higher metallicity do not need any correction. 
A pure spectroscopic \teff\ scale leads to systematically lower abundances for metal-poor stars. 
With the adopted procedure, the \teff\ of all the stars are on the same scale.
On the other hand, for the MW GCs 
homogeneous photometry is available \citep{stetson19} and \teff\ 
have been derived from the ${\rm (V-K_{0})}$-\teff\ calibration \citep{gh09}. 

However, since one of the key results presented in this letter is based on the comparison of chemical 
abundances among LMC and MW clusters, it is particularly important to recall that \teff\ for all 
the cluster stars are on the same scale \citep{gh09}.

The surface gravities have been estimated assuming for each cluster the 
\teff-log~g relation suitable for the red giant branch and derived from a theoretical isochrone \citep{dotter08} 
with an age of 13 Gyr and metallicity and \alfa\ from our chemical analysis. 
Because log~g values have been derived according to the 
\teff\ and metallicity of each star, the procedure to obtain this parameter is iterative.
This approach avoids the uncertainties in log~g arising from 
color excess and distance modulus of each individual cluster. 
The assumption of a different age is not critical: a change of 1 Gyr (that can be consider as a reasonable uncertainty 
in the ages of the target clusters) implies a variation of 0.01 in log~g, with a negligible 
impact on the derived abundances.

Microturbulent velocities are derived 
by requiring no trend between abundances of the iron lines and their strength \citep{m11vt}. 

The 2 observed stars in NGC~2005 have atmospheric parameters comparable with those of the other stars in LMC clusters with similar metallicities, as expected because all the target stars belong to the brightest portion of the clusters red giant branches
(due to their distance, high-resolution spectroscopy in old LMC GCs is restricted to the brightest stars).
We checked that there is no set of reasonable 
atmospheric parameters able to reconcile all the abundances of NGC~2005 with those measured in the other LMC metal-poor clusters. 
Because the analyzed transitions for the 13 measured species have different strengths, excitation potential and ionization stages, 
they have different (and sometimes opposite) sensitivity to the atmospheric parameters. 
Therefore, the variation of a given atmospheric parameter leads 
to an increase of some abundance ratios and the decrease of others, depending on the characteristics of the used transitions.

For example, a decrease of \teff\ by 200 K 
(coupled with new, appropriate log~g and microturbulent velocities) for the stars in NGC~2005 provides [Si/Fe] comparable 
with those of the other clusters while [Ca/Fe] and [Ti/Fe] remain low. Furthermore, this new set of parameters 
provides values of [Zn/Fe] and [Cu/Fe] still significantly lower than those of the other LMC clusters. 
In order to increase these two abundance ratios, \teff\ should be increased by 400-500 K, decreasing 
significantly the other abundance ratios. Also, such hot \teff\ are incompatible with the position of 
the stars on the color-magnitude diagram.
We therefore rule out that the peculiar chemistry of NGC~2005 could be driven by a particular choice 
of the atmospheric parameters of its stars, concluding that its deviations from the average trends defined by the other 
LMC clusters are genuine.

Figure 3 shows some portions of the MIKE spectrum of the star NGC2005-S3, around four metallic lines 
of the species (namely Sc, Zn, La and Eu) that exhibit the largest differences between NGC~2005 and the clusters with similar [Fe/H] 
(see Fig.1). The MIKE spectrum is compared with two synthetic spectra: the best-fit one and the one 
calculated assuming the average abundances measured in the other 5 LMC GCs.
As is evident from this figure, the depth of the observed lines in NGC~2005 is not compatible 
at all with the abundances of the other LMC GCs with similar [Fe/H].
On the other hand, Figure 4 shows the comparison between the MIKE spectrum of the star NGC2005-S3 and 
the UVES spectrum of the star NGC2210-764 (which has has atmospheric parameters and metallicity very similar to those of NGC2005-S3).
A smoothing filter and a re-sampling have been applied to the UVES spectrum to mimic the spectral resolution 
and the pixel-size of MIKE spectra, allowing us to directly compare the line strength of metallic lines. 
As visible in  Figure 5 , Fe lines have similar strengths while Sc, La and Eu lines are shallower 
in the MIKE spectrum of NGC205-S3.

The uncertainty in the line fitting procedure and in the continuum location are not sufficient to justify 
such a stark discrepancy, not even in the case of the Zn line, that is one of the bluest transitions analyzed 
in this study, and for which the continuum location is more affected by the lines crowding.

{\bf Comparison between abundances from UVES and MIKE spectra}\\
NGC~2005 is the only LMC GC among those shown in Fig.1 to have been
observed with MIKE. The other five have been observed with UVES and FLAMES. 
We thus performed some checks on the elemental abundances of NGC~2005 to exclude the possibility that the chemical 
peculiarity of this cluster is artificially caused by some systematic effect due to the use of different spectrographs.

To do so, we performed some checks on the abundances derived from UVES and MIKE. First,
we considered two clusters, namely Hodge 11 and NGC~1898, for which both UVES and MIKE spectra are available 
(though no stars have been simultaneously observed with both the instruments). 
Figure 5 shows the differences between the average abundances as derived from UVES and MIKE spectra 
for these two clusters. No systematic difference exists for any of the measured elemental abundances, 
this demonstrating that the two instruments provide abundances that are fully compatible within the uncertainties.

Similarly to what we did in Fig.1,  Figure 6 compares the average abundance ratios 
measured in two clusters with similar [Fe/H] but observed with the two spectrographs, namely NGC~1754 
(observed with UVES) and NGC~2019 (observed with MIKE). Also for this pair of clusters, no significant differences 
are found and the abundance ratios of NGC~2019 are not systematically lower than those measured in NGC~1754.

As an additional sanity check, we further repeated the analysis of the UVES spectra of Hodge 11 and NGC~1898 by 
applying a smoothing filter, thus to reproduce the spectral resolution of the MIKE spectra, 
and by sampling the spectra to the pixel size of MIKE. This set of {\sl MIKE-like} spectra allows to estimate 
whether some instrumental characteristics of the spectrograph (i.e. spectral resolution, efficiency and pixel size) 
can induce systematic differences in the derived abundances, for instance leading to over- or -under-estimate 
the continuum level or to significant variations in the derivation of the atmospheric parameters 
(temperatures and microturbulent velocities have been derived spectroscopically). 
When these spectra are analyzed after fixing the stellar parameters obtained from the original UVES spectra, 
the average difference between the new and the original iron abundances is -0.03$\pm$0.02 dex ($\sigma$=~0.05 dex). 
When the stellar parameters are re-derived, the average difference in [Fe/H] become -0.05$\pm$0.02 dex ($\sigma$=~0.05 dex), 
due to small changes in the stellar parameters themselves.
In both cases, the characteristics of the MIKE spectra induce only a very small decrease of the iron abundances, 
while the differences cancel out for the [X/Fe] abundance ratios.

Finally, we refer to the recent analysis of the Galactic benchmark star HD20 using both UVES and MIKE 
spectra \citep{hanke20}. Thanks to the very high S/N ratio of the spectra of this bright star ($>$400 for UVES 
and $>$1000 for MIKE), this comparison is adequate to highlight intrinsic differences solely due to the 
instruments (and not induced by the noise). The agreement between the abundances of Ti, Fe and Nd 
(the species with the largest number of available lines in the analysis) derived from the same lines and 
measured with UVES and MIKE is found to be excellent, thus excluding again significant systematics 
between the two instruments.

All these checks demonstrate that the abundances derived from MIKE and UVES are fully consistent with each other 
within the uncertainties and that the low abundance ratios measured in NGC~2005 are not an instrumental artifact.
\\

{\bf Uncertainties in the chemical abundances}\\
The total uncertainty associated to a given abundance (in the form 
of [X/H]) in individual stars is obtained by taking into account internal errors 
(mainly related to the measure methodology) and those arising from the adopted stellar parameters. 
On the other hand, when an abundance ratio [X/Fe]=[X/H]-[Fe/H] is considered, the 
uncertainties arising from atmospheric parameters partially cancel out because 
metallic lines of different species but the same ionization stage respond in a similar way 
to variations in these parameters. 

The uncertainty of [Fe/H] is obtained by summing in quadrature the internal error and the variations 
in the Fe abundance due to errors in atmospheric parameters.
The uncertainty for the abundance ratio [X/Fe] is computed by summing in quadrature the 
internal errors of X and Fe (that are independent each other), and the abundance ratio variation 
due to uncertainties in the adopted atmospheric parameters.

For each cluster, mean abundance ratios (and the corresponding standard errors) have been computed 
by averaging the abundances of the member stars weighted by the uncertainty (as described above).
Since formal standard error on the weighted mean were in many case exceedingly small (of the order of $\sim$0.02--0.03 dex), 
due to the small number of stars per cluster (2-3), we decided to take the average error on individual measures 
as a conservative estimate of the uncertainty on the mean abundance.

{\sl Internal errors ---}
Internal errors in [X/H] were estimated considering the line-to-line dispersion of the abundance mean divided 
by the root mean square of the number of lines. The dispersion of the mean reflects a combination of 
uncertainties in the measure, continuum location and in the atomic data. 

When one only line is available, we considered as internal error the abundance variation due to the uncertainty 
in the measure process. For species for which equivalent width has been measured, we 
 transformed in abundance the error associated to the Gaussian fit used to measure the equivalent width. 
 For species measured from the spectral synthesis, we performed Montecarlo simulations of the fitting procedure. 
For each star, a sample of 500 artificial spectra has been generated, by re-sampling 
the best-fit synthetic spectrum to the instrumental pixel-size and injecting Poissonian noise 
to reproduce the measured signal-to-noise. This sample of artificial spectra has been analyzed 
with the same approach adopted for the real spectra. The dispersion of the derived abundance distribution 
has been adopted as 1$\sigma$ uncertainty.

{\sl Parameters errors ---}
Abundance errors due to uncertainties in the atmospheric parameters were  estimated  by  re-computing abundances 
varying the parameters by their uncertainties. 
The uncertainties in spectroscopic \teff\ are estimated by applying a jackknife bootstrapping technique \citep{lupton93}, 
leading to errors from $\sim$50 up to $\sim$100 K, mainly depending on the signal-to-noise ratio of the spectrum. 
For the clusters with [Fe/H]$<$--1.5 dex, for which the correction to the photometric scale has been applied, 
we added in quadrature also the 1$\sigma$ dispersion (36 K) associated to the calibration itself \citep{mb20}. 
Temperatures were varied by the corresponding errors, gravities were modified by propagating the errors 
in \teff\ on the adopted \teff\--log~g relation and the microturbulent velocities were re-computed adopting 
the new \teff\ and log~g. This approach allows to take into account the covariance existing 
between \teff\ and log~g \citep{cayrel04}, due to the physical relation existing between these two parameters, 
and between \teff\ and microturbulent velocity, due to 
the correlation between line strength and excitation potential.

{\sl Systematic errors ---} 
Chemical abundances can be affected by several sources of systematics, mainly the accuracy of the adopted atomic data, 
the used solar reference abundances, the used model atmospheres (and their physical assumptions), the zero-point 
of the used \teff\ scale, and the method to infer stellar parameters. The chemical analysis of the two datasets discussed 
in this work (LMC and MW GCs) has been performed using the same approach in terms of these assumptions,
in order to erase the main systematics and compare directly the abundances of the two families of clusters. 
Therefore, any possible source of systematic error arising from the analysis affects in the same way both 
the datasets, making the comparison between LMC and MW clusters more accurate and robust.

{\bf Chemical evolution models}\\
    The trends of the abundance ratios of different chemical elements as a function of time (as traced by metallicity) 
    in a given stellar system can be used to infer the structure formation timescale as well as the role of any gas 
    inflow/outflow and the shape of the prevailing IMF. However, in order to do so, one needs to work out the proper 
    chemical evolution model, tailored to the specific object under scrutiny.

   The chemical evolution model for the MW adopted in this paper is described extensively in previous papers \citep{chiappini01,romano10a}. It assumes that the inner Galactic halo forms at early times from the accretion of unprocessed gas that triggers a very efficient star formation, of the order of $\sim$10~M$_\odot$ yr$^{-1}$ on a Gyr timescale. The Galactic disc forms later on at a slower pace, but since our MW GC data trace only the first $\sim$1~Gyr of Galactic evolution, in the following we omit all the details regarding the formation of the disc component -- the interested reader is referred to the original papers. As we will see in the following, it is very important to calibrate the main ingredients of the chemical evolution model against a valid reference template; the MW provides indeed a very good anchor.
   
   The models for the LMC and the putative NGC~2005 parent galaxy rest on previous work for dwarf Galactic satellites \citep{romano13,romano15}. As for the LMC, we implement in the model the global star formation history (SFH) derived from observational pointers independent from chemical indicators (i.e., long-period variable star counts, which agree with previous studies \citep{rezaeikh14}). According to the adopted SFH, most LMC stars (about 75 per cent of the total stellar population) form during the first $\sim$3~Gyrs of evolution. The star formation rate peaks at SFR $\sim$ 1--1.5~M$_\odot$ yr$^{-1}$ during the first 1.5~Gyr of evolution, and steadily declines afterwards. As for the dwarf NGC~2005 progenitor, there are not independent SFH indicators that can be accessed, hence we assume a star formation burst forming $2 \times 10^5$~M$_\odot$ of stars in either 0.5 or 1~Gyr. The star formation of NGC~2005's progenitor galaxy is found to proceed at a much slower pace than that of the LMC, namely, $<$2.5--5~$\times$ 10$^{-4}$~M$_\odot$ yr$^{-1}$ (with the highest values corresponding to the shortest-duration burst).

   In all models, cold gas of primordial chemical composition is accreted at an exponentially decreasing rate:
   \begin{equation}
     \frac{{\rm d}{M}_{\rm inf}(t)}{{\rm d}t} \propto {\rm e}^{-t/\tau}
   \end{equation}
   where ${M}_{\rm inf}(t)$ is the mass accreted at time $t$ and $\tau =$~1, 0.5 and 0.005 Gyr are the e-folding times for the MW, LMC and UFD NGC~2005 progenitor, respectively. 
   We note that this smooth infall law produces results that are in qualitative agreement with those obtained by adopting much more complex accretion histories from cosmological simulations \citep{colavitti08,romano13}.

The star formation rate is implemented according to the Kennicutt-Schmidt law 
\citep{schmidt59,kennicutt98}. In the model for the MW it reads
   \begin{equation}
     \psi(t) \propto \sigma_{\mathrm{gas}}^k(t),
   \end{equation}
   where $\sigma_{\mathrm{gas}} (t)$ is the surface gas density at a given
   time and $k =$~1.5.
   In the models for dwarf galaxies it is
   \begin{equation}
     \psi(t) \propto {M}_{\mathrm{gas}}^k(t),
   \end{equation}
   where ${M}_{\mathrm{gas}}(t)$ is the gas mass at a given time and 
   $k =$~1.

   As for the Galactic halo, the galaxy-wide IMF is the canonical one used in previous work \citep{kroupa02}, 
   with $x =$ 1.7 in the high-mass domain. A slightly steeper galaxy-wide IMF ($x =$ 1.9 in the high-mass domain) 
   is found to fit better the LMC data at relatively high metallicities; therefore, in Fig.~2 we show the results 
   obtained with this IMF choice. Furthermore, in order to reproduce the chemical abundance ratios measured in NGC~2005, 
   we find that it is necessary to require also a reduction of the upper mass limit of the IMF, from 100 to 40~M$_\odot$. 
   These assumptions are justified, at least qualitatively, in the framework of the integrated galactic 
   IMF theory by the low star formation rates and metal-poor environments that characterize dwarf and UFD galaxies \citep{yan20a}. 

{\bf Nucleosynthesis prescriptions}\\
   The most important ingredients of chemical evolution models are the stellar yields, namely, the amounts of different chemical elements that stars produce and eject into the ISM at their deaths. The chemical evolution models adopted in this study track the evolution of the abundances of several elements from hydrogen to europium, allowing us to study the evolution of elements that are produced by various nucleosynthetic processes in stars of different masses and initial chemical composition. The instantaneous recycling approximation is relaxed, i.e. we consider in detail the stellar lifetimes. In this way, different chemical elements are correctly restored to the interstellar medium at different times, according to the lifetimes of their stellar progenitors.

   We adopt grids of stellar yields calibrated against the MW data; in particular, with the adopted prescriptions for single low- and intermediate-mass stars \citep{karakas10}, massive stars \citep{nomoto13} and type Ia SNe \citep{iwamoto99} (thermonuclear explosions of white dwarfs in binary systems \citep{mg86,matteucci01}) we are able to reproduce very well the average trends of the abundance ratios of several elements, including [Si/Fe], [Ca/Fe], [Zn/Fe], and [Cu/Fe], as a function of [Fe/H] in the Galactic halo \citep{romano10a}. In particular, regarding the high-mass stars, we use a mixture of ``normal'' core-collapse SNe, which explode releasing energies of the order of $10^{51}$~ergs, and hypernovae, characterized by much larger explosion energies. In particular, by considering hypernova explosions it is possible to explain the run of [Zn/Fe] with [Fe/H] in halo stars \citep{kobayashi06,romano10a}. We obtain a good fit to the data presented in this paper by assuming that as many as 95 per cent of stars with $m >$~20~M$_\odot$ explode as hypernovae for [Fe/H]$< -2.5$, while the hypernova fraction goes to zero for [Fe/H]$> -1$ dex. In order to fit the MW GC data at best, we further adopt zero-point shifts of $- 0.2$~dex for both [Si/Fe] and [Zn/Fe] (well inside the range allowed by theoretical uncertainties and observational systematics that may affect the ratios).
   The same stellar nucleosynthesis prescriptions (and zero-point shifts) are then adopted in the models for the LMC and NGC~2005's parent galaxy. Interestingly, it is found that the best agreement between model predictions and relevant data is obtained with a galaxy wide IMF that varies in qualitative agreement with the predictions of the integrated galactic IMF theory \citep{yan20a}.

{\bf Do observed counterparts of the progenitor of NGC~2005 exist? }\\
The chemical abundance patterns measured in NGC~2005 and in the other LMC GCs demonstrate that 
the former originated in an environment characterized by a significantly less efficient star formation 
than that of the LMC. This is typical of dwarf spheroidal (dSph) satellites of the Milky Way \citep{tolstoy09}. 
Thus it is natural to search among them, when looking for an existing galaxy similar to the putative progenitor of NGC~2005. 

There are only two dSphs currently orbiting the MW that were able to form globular clusters: Sagittarius and Fornax. 
However, the abundance pattern of the Sgr dSph is very similar to that of the LMC \citep{muc17,minelli21}, 
and as such it is not compatible with the chemical composition of NGC~2005. On the other hand, Fornax seems 
to fit all the properties of the progenitor galaxy of NGC~2005. In fact, the abundance pattern of NGC~2005 
is remarkably similar to that of Fornax stars of the same metallicity.
As we show in Figure 7, other two dSph galaxies, namely Draco and Ursa Minor \citep{shetrone01, letarte06, cohen09, cohen10, letarte10, lemasle14, ural15} 
provide a good chemical match 
when compared to NGC~2005, but they have a stellar mass comparable to NGC~2005 itself \citep[$\simeq 3\cdot 10^5$~M$_{\sun}$,][]{mcconn12}. Instead, Fornax has a stellar mass large enough \citep[$\simeq 2\times 10^7$~M$_{\sun}$,][]{mcconn12}
to host a population of 5 old GCs, four of them being in the same mass range as NGC~2005 \citep[$\ga$1.3$\cdot10^{5} {\rm M_{\odot}}$,][]{leung20}. In general, dwarf galaxies with mass comparable to Fornax typically host between 0 and 6 
globular clusters \citep{prole19}.
The mass ratio between Fornax and LMC is ${\frac{M_{For}}{M_{LMC}}}<0.01$, for both stellar and total (dynamical) mass. 
Therefore, the merging of a progenitor galaxy of NGC~2005 similar to the Fornax dSph with the LMC would classify 
as a minor merger, with negligible consequences on the structure of the LMC and negligible probability 
to leave a long-lived relic, except for a dense cluster with chemical composition not compatible 
with being born in the LMC. 

We conclude that the properties of the hypothesized progenitor galaxy of NGC~2005, now dissolved into the LMC, 
are fully compatible with well known existing galaxies, the Fornax dSph providing the best suited local example.

{\bf Data Availability:} 
Most of the data used in this work are available in the public archive of the 
European Southern Observatory (http://archive.eso.org/eso/eso\_archive\_main.html and 
http://archive.eso.org/wdb/wdb/adp/phase3\_main/form). 
All the data are available from the corresponding author upon reasonable request.

{\bf Code Availability:} 
The codes used for the chemical analysis are publicly available: 
GALA (http://www.cosmic-lab.eu/gala/gala.php), ATLAS9 
(https://wwwuser.oats.inaf.it/castelli/sources/atlas9codes.html), SYNTHE (https://wwwuser.oats.inaf.it/castelli/sources/synthe.html), DAOSPEC (http://www.cadc-ccda.hia-iha.nrc-cnrc.gc.ca/en/community/STETSON/daospec/). 
We opt not to make the code used for the chemical evolution modeling publicly available because it is an important asset of the researchers’ tool-kits.

{\bf Acknowledgements:} 
A.Mucciarelli, FRF and LO acknowledge the financial support of the project "Light-on-Dark" , granted by the Italian MIUR through contract PRIN-2017K7REXT. 
DR gratefully acknowledges the International Space Science Institute (ISSI) in Bern, CH, and the International Space Science Institute-Beijing (ISSI-BJ) in Beijing, CN, for support provided to the team “Chemical abundances in the ISM: the litmus test 
of stellar IMF variations in galaxies across cosmic time". DR and MB acknowledge the financial support of INAF through 
the Main Stream grant CRA 1.05.01.86.28 assigned to the project “SSH: the Smallest Scale of Hierarchy”.
A.Mucciarelli thanks Jennifer Johnson for sharing the MIKE spectra.
A.Mucciarelli dedicates this work to the memory of his twin brother Simone.

{\bf Authors contributions:} 
A.Mucciarelli designed the study, coordinated the work and performed the data analysis, DM led the scientific interpretation, 
A.Minelli contributed to the spectroscopic analysis, DR computed the chemical evolution models and contributed to the scientific 
interpretation, MB contributed to the scientific interpretation and to the writing, FRF, FM and LO contributed 
to the presentation of the paper. All the authors critically contributed to the work presented here.

{\bf Competing interests:} 
The authors declare that they have no competing financial interests.

%
%



%
%
\newpage

\begin{figure}
 \centering%
\includegraphics[scale=0.37]{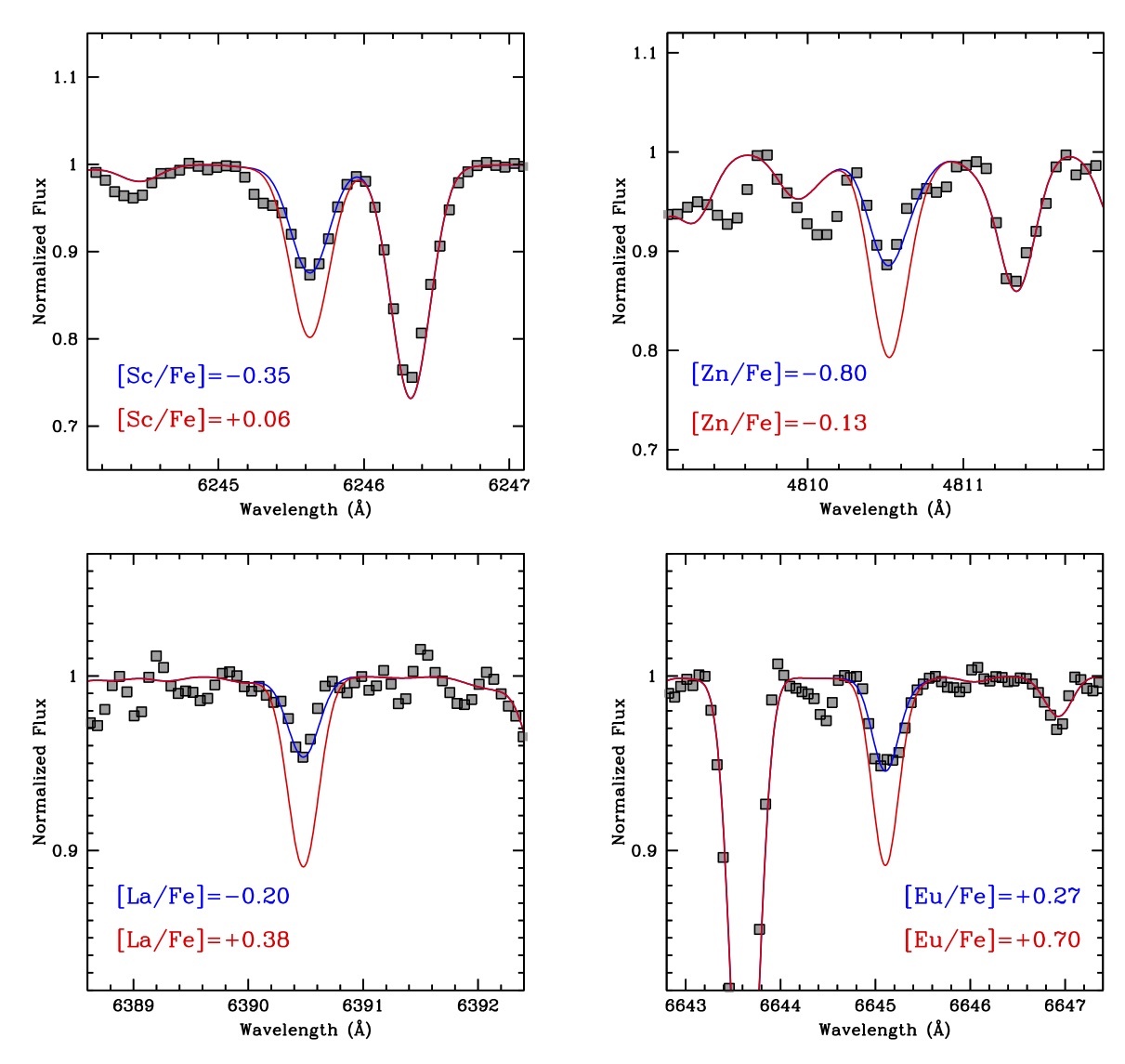}
\caption{{\bf The observed spectrum of NGC2005-S3}. Portions of the MIKE spectrum of the star NGC2005-S3 (gray squares) around some metallic lines of interest for Sc, Zn, La and Eu, with superimposed the best-fit synthetic spectrum (blue lines) and a synthetic spectrum computed with the stellar parameters of this star but assuming the
average abundances derived from the 5 LMC GCs with metallicity comparable to that of NGC~2005 (red lines).}
\label{check_ss}
\end{figure}

\begin{figure}
 \centering%
\includegraphics[scale=0.9]{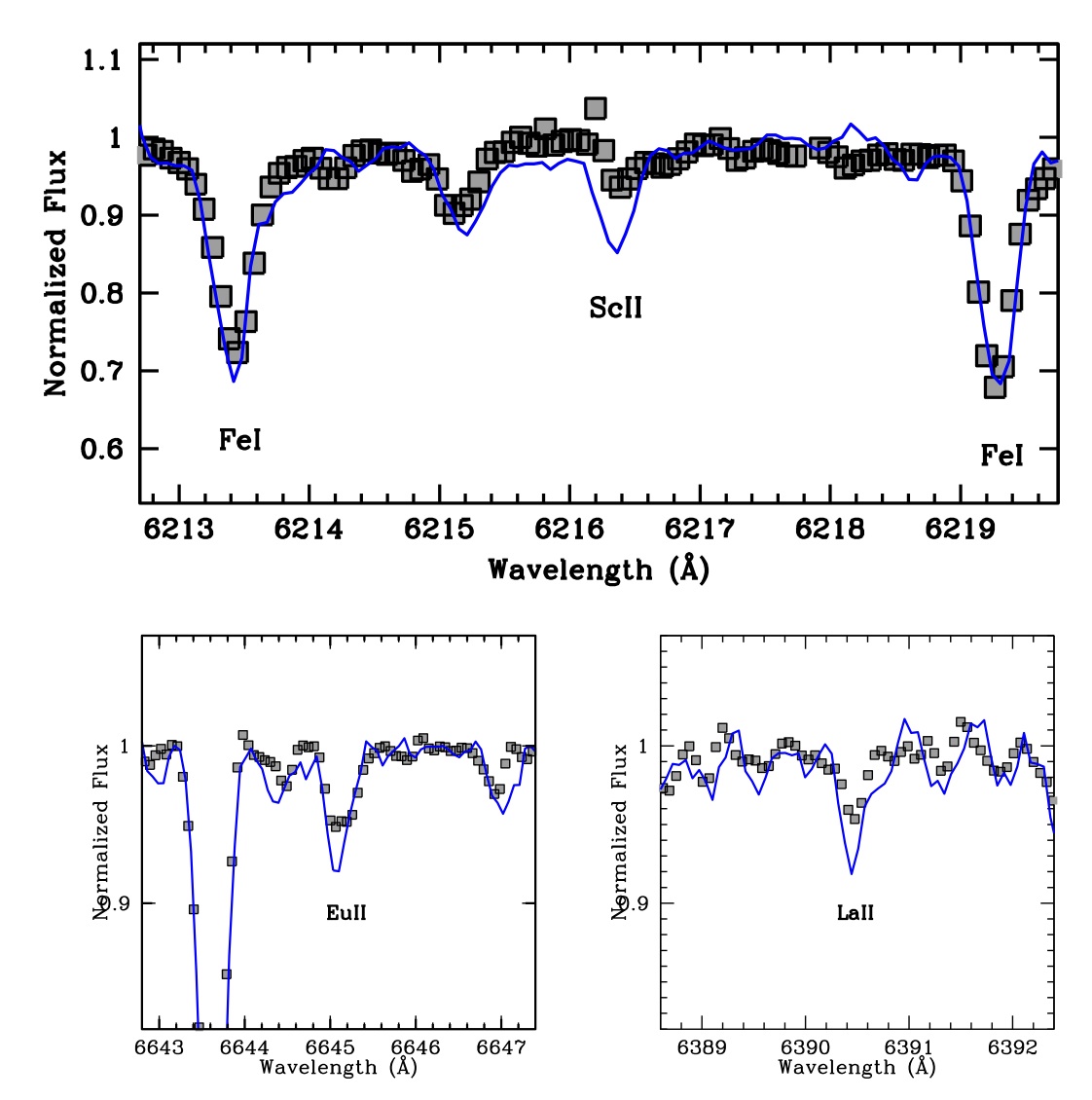}
\caption{
{\bf The observed spectrum of NGC2005-S3}. Portions of the MIKE spectrum of the star NGC2005-S3 
(gray squares) around some metallic lines of interest for Sc, Fe, La and Eu, with superimposed the UVES spectrum of the star NGC2210-764, 
convoluted with a Gaussian profile to reproduce the MIKE spectral resolution and re-sampled to the MIKE pixel size (blue line).}
\label{like}
\end{figure}

\begin{figure}
 \centering%
\includegraphics[scale=0.37]{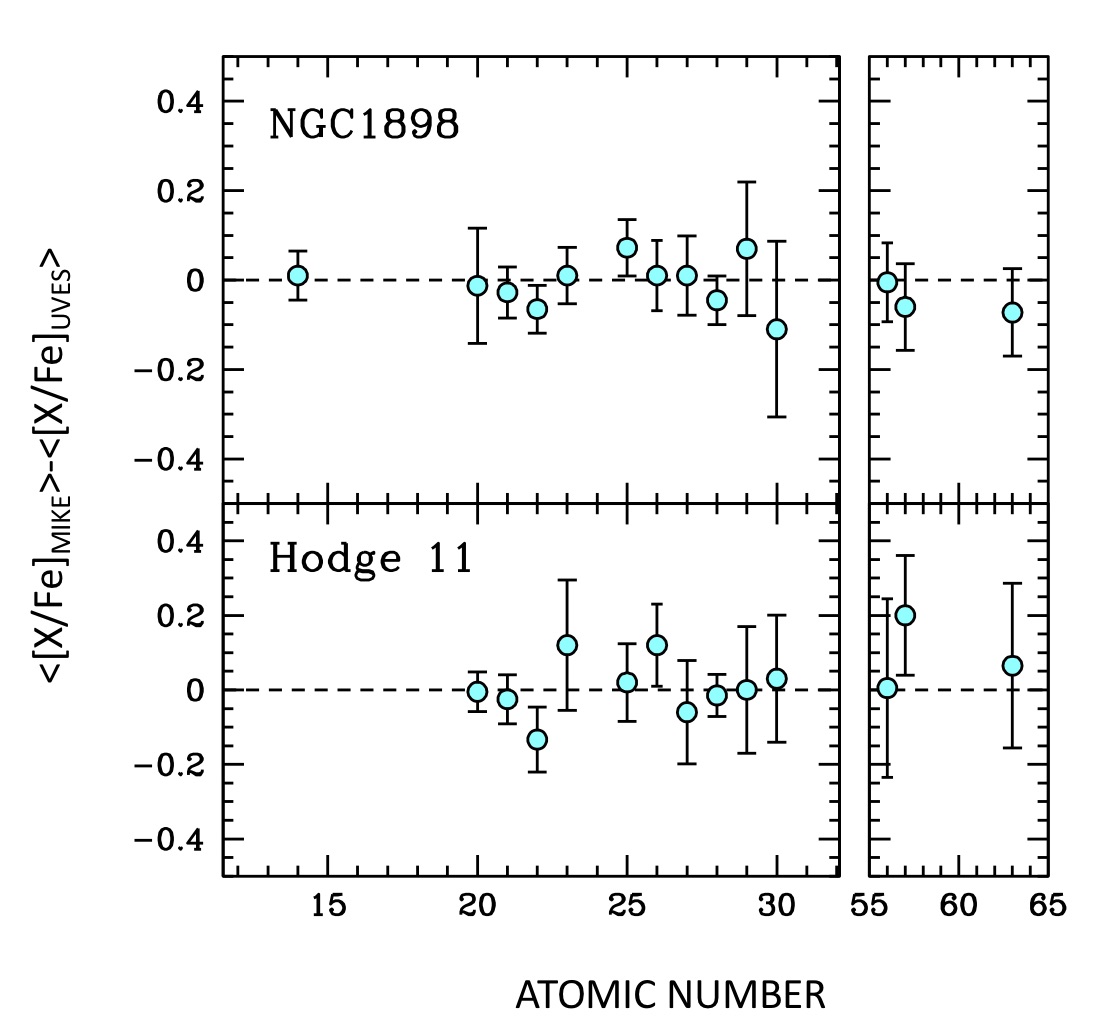}
\caption{
{\bf Comparison between the chemical abundances of NGC~1898 and Hodge 11}. Differences between the average abundances derived from UVES and MIKE spectra for 
the clusters NGC~1898 (upper panel) and Hodge 11 (lower panel).}
\label{comp_um}
\end{figure}

\begin{figure}
 \centering%
\includegraphics[scale=0.6]{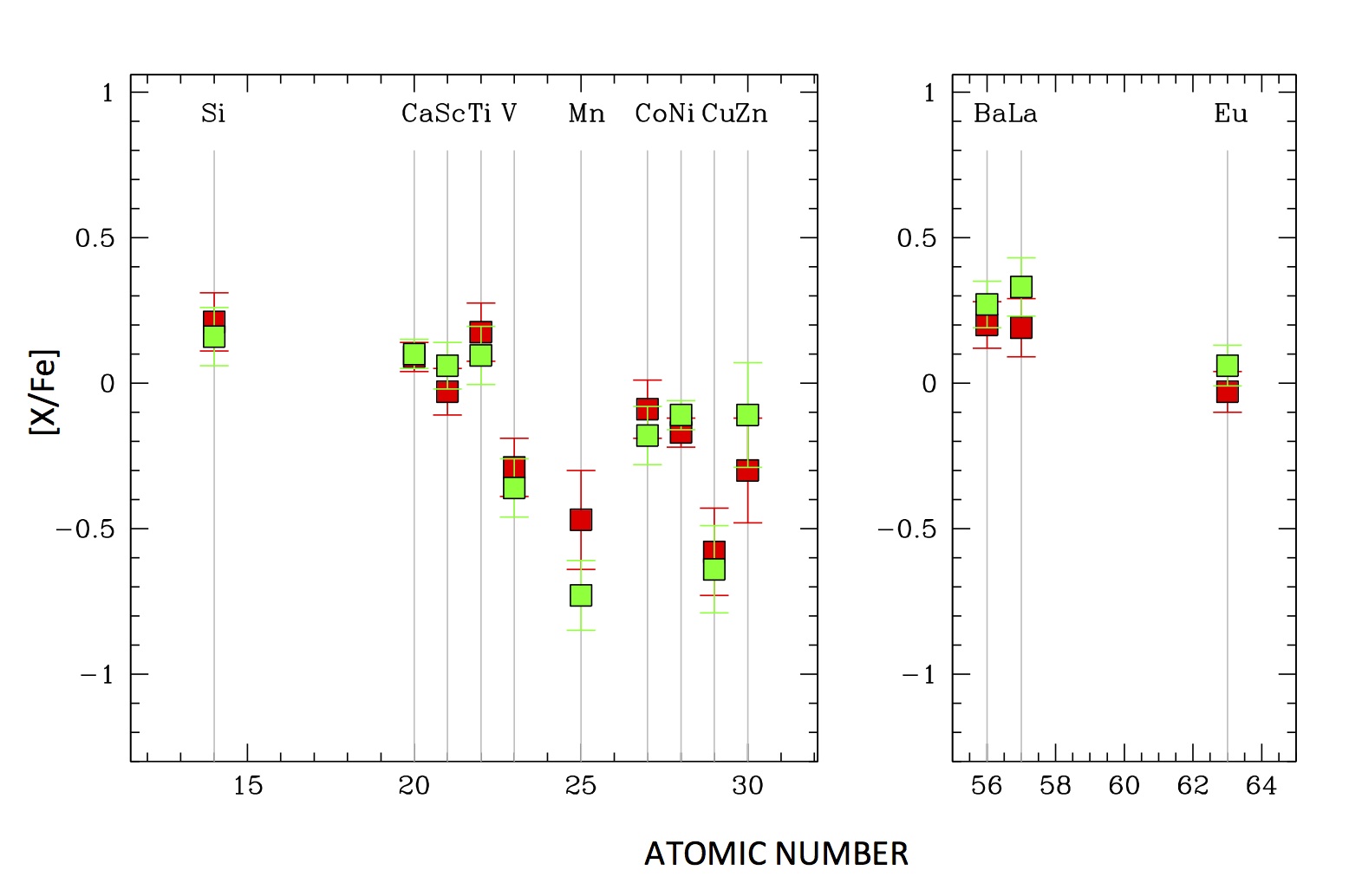}
\caption{{\bf Comparison between the chemical abundances of NGC~1754 and NGC~2019}. Abundance ratios measured for the clusters NGC~2019 (measured with MIKE, red square) 
and NGC~1754 (measured with UVES, green square).}
\label{comp_um2}
\end{figure}

\begin{figure}
 \centering%
\includegraphics[scale=0.28]{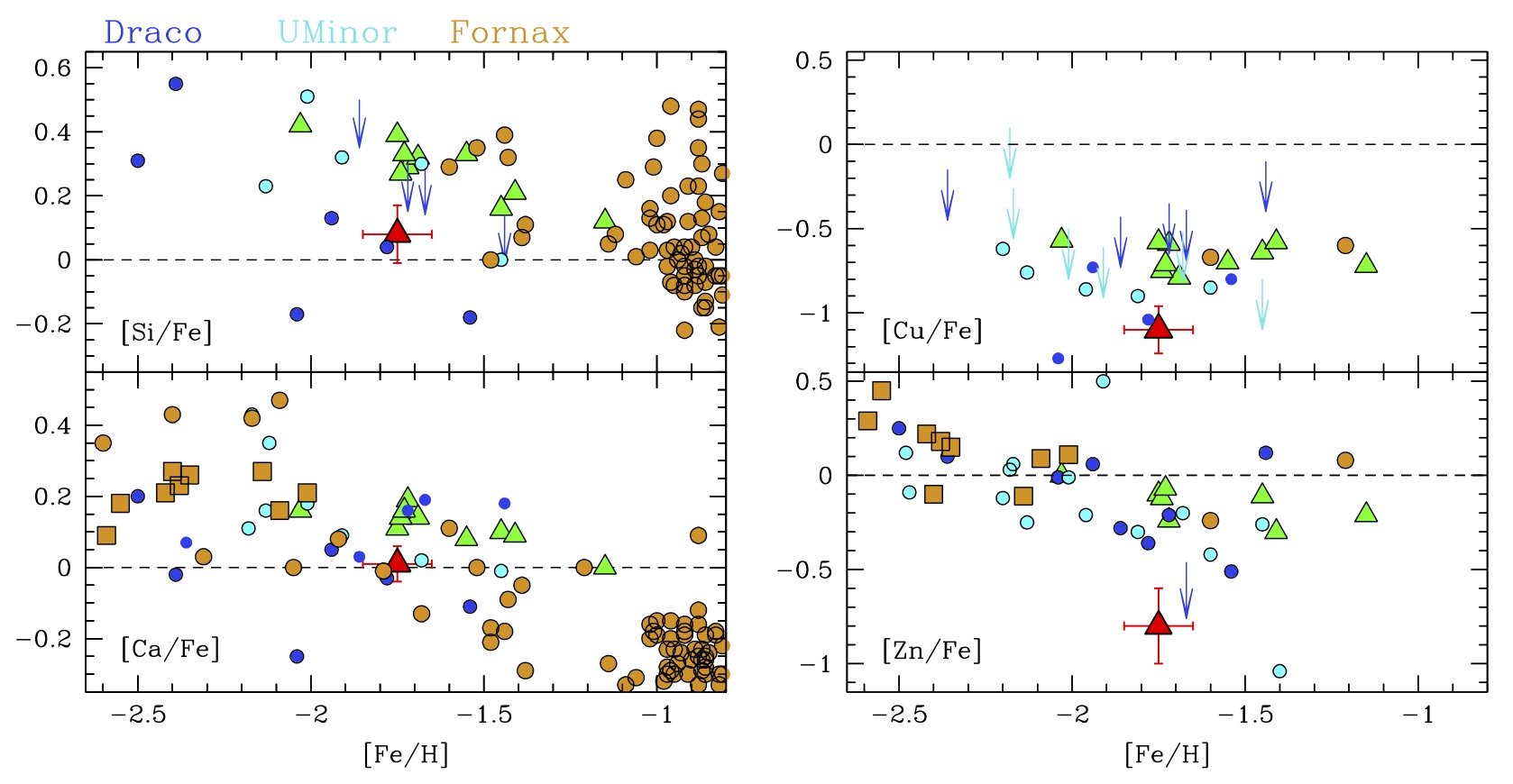}
\caption{{\bf Chemical abundances of the LMC and MW clusters, and dwarf galaxies stars}. 
Behaviour of the [Si/Fe], [Ca/Fe], 
[Cu/Fe] and [Zn/Fe]
abundance ratios as a function of [Fe/H] 
for the accreted 
LMC cluster NGC~2005 (red triangle) and the LMC clusters (green triangles) and 
the field stars in the dwarf spheroidal galaxies Fornax, 
Draco and Ursa Minor (orange, blue and cyan points, 
respectively; arrows indicate upper limits) and individual stars in the Fornax clusters
\citep[orange squares,][]{shetrone01,letarte06,cohen09,cohen10,letarte10,lemasle14,ural15}.}
\label{dwarf}
\end{figure}

\newpage

\begin{deluxetable}{lccccccc}
\tablecolumns{8} 
\tablewidth{0pc}  
\tablecaption{LMC globular clusters dataset}
\tablehead{ 
\colhead{Cluster} &   RA &  Dec & ${\rm N_{U-FL}}$ & ${\rm N_{U}}$ & ${\rm N_{G }}$ &
${\rm N_{M}}$ & Programs \\
  &   (J2000)  &  (J2000)  &  & &  &  & }
\startdata 
\hline
 NGC~1466       &   03:44:33.0  & --71:40:18.0    &  5  &  ---   &  4   &  --- &   092.D-0244  \\
 NGC~1754       &   04:54:18.1  & --70:26:32.6    &  5  &  ---   & ---  &  --- &   084.D-0933  \\
 NGC~1786       &   04:59:07.5  & --67:44:45.0    &  4  &  ---   &  3   &  --- &   080.D-0933  \\
 NGC~1835       &   05:05:09.2  & --69:24:21.0    &  4  &  ---   & ---  &  --- &   092.D-0244  \\
 NGC~1898       &   05:16:45.4  & --69:39:16.7    &  4  &  ---   &  3   &  2   &   084.D-0933 + J06  \\
 NGC~1916       &   05:18:37.9  & --69:24:22.9    &  4  &  ---   &  --- &  --- &   092.D-0244  \\
 NGC~2005       &   05:30:08.5  & --69:45:14.4    &  ---&  ---   &  --- &  2   &   J06  \\
 NGC~2019       &   05:31:56.5  & --70:09:32.5    &  ---&  ---   &  --- &  3   &   J06  \\
 NGC~2210       &   06:11:31.3  & --69:07:17.0    &  5  &  3     &  --- &  --- &   080.D-0368 + UVES-SV  \\
 NGC~2257       &   06:30:12.0  & --64:19:36.0    &  3  &  3     &  3   &  --- &   080.D-0368, 66.B-0331  \\
 HODGE~11       &   06:14:22.9  & --69:50:54.9    &  4  &  ---   &  --- &  2   &   082.B-0458 + J06  \\
 \hline
\hline
\enddata 
\tablecomments{$~~~~~$ 
Coordinates of the cluster centers are from the SIMBAD database. 
The number of analysed member stars for each cluster is listed 
according to the used instruments: U-FL for UVES-FLAMES, U for UVES, 
G for GIRAFFE/MEDUSA-FLAMES, M for MIKE. The program identification 
numbers of the ESO Programs are reported 
(Program ID: 080.D-0368, PI: Origlia; 
Program ID: 084.D-0933, PI: Mucciarelli; 
Program ID: 092.D-0244, PI: Mucciarelli). 
The clusters observed with the spectrograph MIKE are labeled as J06\citep{johnson06}. UVES-SV identifies observations performed during the UVES Science Verification.}
\end{deluxetable}

\begin{deluxetable}{lcccc}
\tablecolumns{5} 
\tablewidth{0pc}  
\tablecaption{MW globular clusters dataset}
\tablehead{ Cluster & RA  &  Dec & ${\rm N_{stars}}$ & Programs  \\
  &   (J2000)  &  (J2000)  &  & }
\startdata 
\hline\hline                       

  NGC~104   & 00 24 05.67     &  --72 04 52.6       & 10    &   073.D-0211 \\   
  NGC~288   & 00 52 45.24     &  --26 34 57.4       & 10    &   073.D-0211 \\   
  NGC~1851  & 05 14 06.76     &  --40 02 47.6       & 23    &   188.B-3002 \\   
  NGC~1904  & 05 24 11.09     &  --24 31 29.0       & 10    &   072.D-0507 \\   
  NGC~2808  & 09 12 03.10     &  --64 51 48.6       & 12    &   072.D-0507 \\   
  NGC~4590  & 12 39 27.98     &  --26 44 38.6       & 13    &   073.D-0211 \\  
  NGC~5634  & 14 29 37.23     &  --05 58 35.1       &  7    &   093.B-0583  \\  
  NGC~5824  & 15 03 58.63     &  --33 04 05.6       &  6    &   095.D-0290 \\   
  NGC~5904  & 15 18 33.22     &   +02 04 51.7       & 14    &   073.D-0211 \\   
  NGC~6093  & 16 17 02.41     &  --22 58 33.9       &  9    &   083.D-0208   \\ 
  NGC~6397  & 17 40 42.09     &  --53 40 27.6       & 12    &   073.D-0211   \\ 
  NGC~6752  & 19 10 52.11     &  --59 59 04.4       & 12    &   073.D-0211 \\   
  NGC~6809  & 19 39 59.71     &  --30 57 53.1       & 13    &   073.D-0211 \\   
  NGC~7078  & 21 29 58.33     &   +12 10 01.2       & 13    &   073.D-0211 \\  
  NGC~7099  & 21 40 22.12     &  --23 10 47.5       & 19    &   073.D-0211 ; 085.D-0375  \\

\hline
\enddata 
\tablecomments{$~~~~~$ Coordinates of the cluster centers are from the Harris catalog.
The number of used stars and the identification numbers of the corresponding ESO Programs are also listed. }

\end{deluxetable}

\end{document}